\documentclass[a4paper,11pt]{article}
\pdfoutput=1
\usepackage{jcappub}
\usepackage{graphicx}
\usepackage{url}

\usepackage[T1]{fontenc}
\usepackage[utf8]{inputenc}
\usepackage[english]{babel}
\usepackage{threeparttable}
\usepackage{amsmath}
\usepackage{amssymb}
\usepackage[toc,page]{appendix}
\usepackage{bm}
\usepackage{braket}
\usepackage{booktabs}
\usepackage{comment}
\usepackage{csquotes}
\usepackage{bookmark}
\usepackage{mathrsfs}
\usepackage{mathtools}
\usepackage{pgfplots}
\usepackage{subfig}
\usepackage{cleveref}

\newcommand{\al}{\alpha}
\newcommand{\be}{\beta}
\newcommand{\ga}{\gamma}

\newcommand{\De}{\Delta}

\renewcommand{\th}{\theta}

\newcommand{\La}{\Lambda}

\newcommand{\m}{\mu}
\newcommand{\n}{\nu}

\newcommand{\vp}{\varphi}

\renewcommand{\c}{\chi}

\renewcommand{\O}{\Omega}
\renewcommand{\d}{\text{d}}

\DeclarePairedDelimiter{\abs}{\lvert}{\rvert}

\newcommand{\fchi}{f\left(\chi \right)}
\newcommand{\phidot}{\dot{\phi}}
\newcommand{\rdot}{\dot{r}}

\newcommand{\beq}{\begin{equation}}
\newcommand{\eeq}{\end{equation}}

{\left\lbrace\begin{array}{@{}l@{}}}%
{\end{array}\right.}

\begin{document}

\title{\boldmath Strong gravitational lensing in $\fchi=\chi^{3/2}$ gravity}


\author[a,b,1]{M. C. Campigotto,\note{Corresponding author.}}
\author[a,b]{A. Diaferio,}
\author[c]{X.Hernandez}
\author[d,b]{and L. Fatibene}


\affiliation[a]{Dipartimento di Fisica, Università degli Studi di Torino,\\Via P. Giuria 1, 10125, Torino, Italy}
\affiliation[b]{Istituto Nazionale di Fisica Nucleare (INFN), sezione di Torino,\\Via P. Giuria 1, 10125, Torino, Italy}
\affiliation[c]{Instituto de Astronomia, Universidad Nacional Autonoma de Mexico, Ciudad de Mexico 04510, Mexico}
\affiliation[d]{Dipartimento di Matematica, Università degli Studi di Torino,\\Via C. Alberto 10, 10123, Torino, Italy}

\emailAdd{martacostanza.campigotto@to.infn.it}
\emailAdd{diaferio@ph.unito.it}
\emailAdd{xavier@astro.unam.mx}
\emailAdd{lorenzo.fatibene@unito.it}

\abstract{
We discuss the phenomenology of gravitational lensing in the purely metric $\fchi$ gravity, an $f(R)$ gravity where
the action of the gravitational field depends on the source mass. We focus on the strong lensing regime in galaxy-galaxy lens systems and in clusters of galaxies. Using an approximate metric solution accurate to second order of the velocity field $v/c$, we show how, in the $\fchi=\chi^{3/2}$ gravity, the same light deflection can be produced by point-like lenses with masses smaller than in General Relativity; this mass difference increases with increasing impact parameter and decreasing lens mass. However, for sufficiently massive point-like lenses and small impact parameters, $\fchi=\chi^{3/2}$ and GR yield indistinguishable light deflection angles: this regime occurs both in observed galaxy-galaxy lens systems and in the central regions of galaxy clusters. In the former systems, the GR and $\fchi$ masses are compatible with the mass of standard stellar populations 
and little or no dark matter, whereas, on the scales of the core of galaxy clusters, the presence of substantial dark matter is required both in General Relativity, and in our approximate $\fchi=\chi^{3/2}$ point-like lens solution.
We thus conclude that our approximate metric solution of $\fchi=\chi^{3/2}$ is unable to describe the observed phenomenology of the strong lensing regime without the aid of dark matter.


}

\maketitle

\section{Introduction}
\label{sec:intro}
Since the first discrepancies between luminous and kinematical masses were observed \cite{Zwicky1933, Smith1936, Zwicky1937}, two different and parallel approaches were developed for their explaination: either these discrepancies are the sign of the existence of some form of dark matter made of so-far hypothetical particles belonging to a wide series of suggested candidates \cite{Bertone2005,Sanders2010}; or they suggest the necessity of modifying General Relativity (GR) on scales larger than the Solar system, where GR is known to describe the observed phenomenology to high accuracy \cite{Diaferio2008}.

In addition to the existence of dark matter to explain the dynamics of cosmic structures, the assumption of the validity of GR implies the existence of dark energy required to describe the accelerated expansion of the Universe when described by a Friedmann model (see e.g. \cite{Weinberg2013} for a review on observational probes of cosmic acceleration). These assumptions are the backbone of the currently most accepted $\La$-Cold Dark Matter ($\La$CDM) cosmological model. $\La$CDM reproduces a wide number of large-scale observations exquisitely well, from the expansion history of the Universe to the power spectrum of the temperature fluctuations in the cosmic microwave background radiation and the scale of the baryonic acoustic oscillations \cite{PlanckCosmo2015}.

However, on galactic scales some challenges arise: the missing satellite problem, the too-big-to-fail problem, the angular momentum catastrophe and the cusp-core problem (see, e.g., \cite{DelPopolo2013} for a review). In addition, the baryonic Tully-Fisher relation in disk galaxies \cite{McGaugh2005}, between the total baryonic mass and the asymptotic rotational velocity, $M_b \sim v^4$, appears to have a scatter of $\sim 0.1$~dex \cite{Lelli2016}, which is slightly smaller than the $\sim 0.15$~dex expected in the $\La$CDM model \cite{Dutton2012}.

These $\La$CDM tensions on small scales vanish when galactic dynamics, rather than being described by GR and dominated by dark matter, is described  by  Modified Newtonian Dynamics (MoND) (see e.g. \cite{FamaeyMcGaugh2012} for a review), which modifies the gravitational force when the acceleration drops below a critical value $a_0$ \cite{Milgrom1983}: for $a\gg a_0$ the force law recovers Newtonian gravity $a \simeq a_N$, while for $a \ll a_0$ the force law is modified such as $a \simeq \sqrt{a_N a_0}$.
MoND fits the rotation curves of a large number of disk galaxies with remarkable accuracy \cite{SandersMcGaugh2002,Gentile2011,Hees2016}, although, for this agreement to hold, it seems that some galaxies should be a factor of two thinner than observed \cite{Angus2015}. 

On larger scales, MoND fails at describing the dynamics of galaxy clusters without assuming the existence of some form of dark matter (e.g. \cite{Aguirre2001,FamaeyMcGaugh2012}). In addition MoND  is a non-relativistic theory of gravity and, by construction, can not describe cosmological phenomena or gravitational lensing, for example the lensing maps of structures like the ``Bullet Cluster'', which is considered the definitive proof of the existence of dark matter \cite{Clowe2006}.  Hence, finding a relativistic extension of MoND is a requirement to assess its viability in the non-relativistic regime. 

A popular attempt is TeVeS \cite{Bekenstein2004,Sanders2005}, which correctly reduces to MoND in the very weak field limit and requires the addition of tensor, vector and scalar fields, that actually mimic the gravitational pull of the dark matter present in the standard model. Unfortunately, some serious tensions appear in TeVeS: for example, this theory is unable to simultaneously reproduce galactic rotation curves and strong lensing data \cite{Ferreras2009}; in addition, a multicentred system can generate a weak lensing signal resembling that of merging galaxy clusters with a bullet-like light distribution but only with the contribution of some dark matter composed of neutrinos of $\sim 2$~eV mass  \cite{Angus2006,Angus2007}, that is the upper bound of current experimental investigations.

A different approach to investigate the relativistic extension of MoND is to explore the $f(R)$ theories of gravity, with $R$ the Ricci scalar, which are the simplest relativistic modification of GR and can satisfy both cosmic and Solar system constraints \cite{DeFelice2010}. However, unlike the standard $f(R)$ theories, the fact that MoND is an acceleration-based modification of gravity and that the Christoffel symbol (the relativistic analogue of the acceleration) is not a tensor requires the introduction of an acceleration scale in the Lagrangian density together with the Ricci scalar.  

In 2011, Bernal et al.~\cite{Bernal2011} propose the metric theory of gravity $\fchi=\chi^{3/2}$, where $\chi$ is a dimensionless Ricci scalar, i.e.~$R$ times a length scale containing the MoND acceleration scale $a_0$. The $\fchi$ theory reduces to a MoNDian-like theory in the non-relativistic limit. They have shown that the theory accounts for the rotation curves and the baryonic Tully-Fisher relation of disk galaxies and the dynamics of X-ray galaxy clusters without any need for dark matter. Bernal et al.~\cite{Bernal2015} show that a perturbed solution for the MoNDian regime  at the fourth order can reproduce the observed temperature profiles of 12 galaxy clusters from Chandra X-ray data without the need of dark matter. Similarly, Mendoza et al. \cite{Mendoza2013} derive a point-like lens solution accurate to second order in $v/c$, consistent with the baryonic Tully-Fisher relation, that yields the same bending angle as a singular isothermal sphere in GR; this solution can be used to investigate the gravitational lensing phenomena, although Mendoza et al. do not test their results against any observed lensing systems.

Given the promising results of this theory derived in the work mentioned above, here we explore the viability of the $\fchi$ theory in more detail by comparing its predictions with gravitational lensing data. We adopt the simplest approach of estimating the expected light deflection magnitude for galaxy-like and cluster-like parameters by assuming $\fchi$ valid and a complete absence of dark matter. Then we compare our theoretical predictions with a sample of galaxy-galaxy lens systems from the SLACS and BELLS catalogues \cite{SLACSIX2009,BELLS2012} and with a limited sample of known galaxy clusters that exhibit strong lensing arclets.

In Section \ref{sec:fchireview} we review the basics of the $\fchi$ theory in the weak field limit, i.e.~$v\ll c$, and present a static and spherically symmetric metric solution of the vacuum. We provide an analytical expression for the deflection angle in the $\fchi$ theory at the order $\mathcal{O}(v^2/c^2)$ for point-like mass lenses. In Section \ref{sec:deflection} we show how the deflection angle depends on the lens mass and the impact parameters in the framework of the new theory and in GR. Finally, in Section \ref{sec:ringssample} we compare our results with real lens systems, both at the galactic and galaxy cluster scales. Contrary to the expectations, our results show that this vacuum solution, approximate to second order in $v/c$, of
the power-law $\fchi$ theory is unable to reproduce the observed deflection angle in the strong lensing regime on the cluster scales without invoking the same amount of dark matter that GR requires.


\section{Basics of $\fchi$ theory of gravity}
\label{sec:fchireview}

In the purely metric $\fchi$ theory, $\fchi$ is an arbitrary analytical function of 
\beq
\c = L^2_M R,
\eeq
where $R$ is the standard Ricci scalar and
\beq
L_M=\zeta r_s^{1/2} R_M^{1/2}\; ;
\eeq
here  $\zeta$ is a coupling constant, of the order of a few, that can be fixed according to Palatini formalism \cite{Mendoza2016};
\beq
r_s=\frac{G M}{c^2}, \qquad R_M = \left( \frac{G M}{a_0}\right)^{1/2}
\label{mondradius}
\eeq
are two length scales that both depend on fundamental constants and on the mass $M$ of the system whose gravitational dynamics we wish to describe. The constant $a_0$ is Milgrom's acceleration whose value we assume to be $a_0=1.2 \times 10^{-10}\ \text{m~s}^{-2}$ \cite{Milgrom1983}.

Therefore, unlike the standard approach where the action of the gravitational field is independent of the system, here the corresponding action 
\beq
S=-\frac{c^3}{16 \pi G L_M^2}\int \d^4x \sqrt{-g} \fchi
\label{actionchi}
\eeq
depends on the system mass, unless $\fchi=\chi$, which corresponds to the Hilbert-Einstein action. For the sake of simplicity, we consider the generic function $f$ to be a power-law 
\beq
f(\c)=\c^b.
\label{powerlaw}
\eeq 
Bernal et al.~\cite{Bernal2011} and Mendoza \cite{Mendoza2012} have shown that the function $\fchi$ recovers a MoND-like acceleration when $b=3/2$. As MoND theory recovers Newtonian theory at certain scales of acceleration, also $\fchi$ theory should reproduce GR at the appropriate scale. Then the function $\fchi$ must satisfy
\beq
    \fchi=
\begin{dcases}
    \c,& \text{when } \c \gg \c_L\\
    \c^{3/2},& \text{when } \c \ll \c_L .
\end{dcases}
\label{systemchi0}
\eeq
for some curvature scale $\chi_L$, that we specify below.

The field equations obtained from the variation of the action \eqref{actionchi} with respect to both $R$ and $M$ are fourth order equations as in standard $f(R)$ theories \cite{Bernal2011}. Mendoza et al. \cite{Mendoza2013} and Bernal et al. \cite{Bernal2015} present a perturbation analysis in the weak field limit to second and fourth order of the velocity field $v/c$. The perturbation to fourth order is required when dealing with velocity dispersions in galaxy clusters, that are typically of order $10^{-4}-10^{-3}$ times the speed of light $c$ \cite{Bernal2015}.
When dealing with gravitational lensing effects, we can limit our analysis to the second order metric solution both in the time and radial components, to obtain bending angle solutions to the same order.
Thus, for a static and spherically symmetric spacetime, the line element of the vacuum solution is
\beq
ds^2=g_{\m\n} dx^\m dx^\n =g_{00}(r) c^2 dt^2 + g_{11}(r) dr^2 - r^2 d\O^2\; ,
\eeq
where $g_{00}$ and $g_{11}$ only depend on the radial coordinate $r$, and $d\O$ is the solid angle element. 

Mendoza et al.~\cite{Mendoza2013} derive the time and the radial components of the second order solution of the $\fchi$ theory by combining the observed baryonic Tully-Fisher relation with the requirement that the deflection angle in 
gravitational lensing is independent of the impact parameter, as suggested by observations. Following Mendoza et al.~\cite{Mendoza2013}, we find
\begin{align}
g_{00}&=1+g_{00}^{(2)} + \mathcal{O}(4) \label{metricg00a} \\ 
g_{11}&=-1-g_{11}^{(2)} + \mathcal{O}(4) \label{metricg11a}
\end{align}
where
\begin{align}
g_{00}^{(2)}&=\frac{2(GM a_0)^{1/2}}{c^2} \log \left(\frac{r}{r_*}\right) \label{metricg00b}\\
g_{11}^{(2)}&=\frac{2(GM a_0)^{1/2}}{c^2} \label{metricg11b}
\end{align}
with $M$ the total mass of the system only assumed to be baryonic, and $r_*$ a scale radius; $r_*$ is a free parameter, because the metric components only  determine the gravitational potential, whose zero point is unconstrained.
Here, we assume $r_*=R_M=(G M / a_0)^{1/2}$, with $R_M$ the natural length scale of the system in the $\fchi$ theory.
We will show below that the resulting deflection angle basically is independent of the choice of $r_*$.

Our metric solution \eqref{metricg00b} has the opposite sign of the same term $ g_{00}^{(2)}$ reported in equation (59) of \cite{Mendoza2013}. 
However, the sign choice is not arbitrary, because it yields the sign of the gravitational potential and it does not depend on the chosen signature of the metric. In Appendix \ref{appendixA},  we present a formal derivation of equation \eqref{metricg00b}.
Thus, in the weak field limit, the gravitational potential corresponding to the metric \eqref{metricg00a}-\eqref{metricg11b} is given by
\beq
g_{00} \simeq 1+\frac{2\phi}{c^2}
\eeq
where
\beq
\phi(r)=v^{2} \log \left(\frac{r}{r_*}\right)=(GM_ba_0)^{1/2} \log \left(\frac{r}{r_*}\right)
\label{wrongpot}
\eeq
which yields the radial acceleration $a= - \left( G M_b a_0\right)^{1/2}/r$ and the baryonic Tully-Fisher empirical relation $GM_b= v^4/a_0$.

With this explicit expression of the metric solution, we can compute the curvature scale $\c_L$ in the Lagrangian density \eqref{systemchi0}. The Ricci scalar corresponding to the metric of \cref{metricg00a,metricg11b} is
\begin{equation}
R = \frac{2 \ga \left[8 \ga^2 \log ^2\left(\frac{r}{R_m}\right)+6 \ga \log \left(\frac{r}{R_m}\right)+\ga+1\right]}{(2 \ga+1) \left[2 \ga
   r \log \left(\frac{r}{R_m}\right)+r\right]^2}
\label{ricci}
\end{equation}
where $\ga=\frac{\left( G M a_0\right)^{1/2}}{c^2}=(v/c)^2$.
The Ricci scalar, to first order in $\ga$, becomes
\beq
R \approx \frac{2 \ga}{r^2} +\mathcal{O}\left(\ga^2\right) \; .
\eeq
Because we want to have the transition scale between $\chi^{3/2}$ and GR at the length scale $R_M$, we  
impose $r=R_M$ in the above equation, and we obtain the curvature scale of the transition  
\beq
\chi_L=\zeta^2 r_s R_m \frac{2 \ga}{R_m^2}=2 \zeta^2 \left( \frac{r_s}{R_m} \right)^2 \; .
\eeq
Neglecting the constant $2 \zeta^2$, we can thus finally define the function $\fchi$ 
\beq
    \fchi=
\begin{dcases}
    \c,& \text{when } \c \gg \chi_L\sim \left( \frac{r_s}{R_m} \right)^2\\
    \c^{3/2},& \text{when } \c \ll  \chi_L\sim \left( \frac{r_s}{R_m} \right)^2 .
\end{dcases}
\label{systemchi}
\eeq


\section{Deflection angle in the $\mathbf{\fchi}$ theory}
\label{sec:deflection}
In this section, we review formulas and relations for the deflection angle to order $(v/c)^2$ for a point-mass lens, and
we show how the deflection angle depends on the lens mass and on the impact parameter at galactic and cluster 
scales.

Under the action of a gravitational field, the total deflection angle of a light ray in an asymptotically flat, spherically symmetric spacetime is given by \cite{Weinberg}
\beq
\De\vp=2\int_{r_0}^{\infty} \! \be \ dr  -\pi
\label{dephi}
\eeq
where
\beq
\be= \frac{\left[-g_{00}(r)g_{11}(r)\right]^{1/2}}{r\left[(r/r_0)^2 g_{00}(r_0) - g_{00}(r) \right]^{1/2}}
\label{betacomplete}
\eeq
and $r_0$ is the distance of closest approach to the lens, which is related to the impact parameter $b$ through the relation $r_0^2=b^2 g_{00}(r_0)$ \cite{Pireaux1999}. In the weak field limit, corresponding to $2GM/c^2\ll r$ in GR and $GMa_0/c^4\ll 1$ in $\fchi$, we can use the approximation $b \approx r_0$. Equation \eqref{dephi} holds only for an asymptotically flat spacetime where the light ray is a straight line in the absence of a 
lens.
In a non-flat spacetime, the identity $r_0^2=b^2 g_{00}(r_0)$ does not hold, since both the impact parameter and $\De\vp$ itself \eqref{dephi} are not well defined \cite{CampigottoCG2}. Rindler and Ishak \cite{RindlerIshak2007} proposed an alternative approach to define the deflection angle in the neighborhood of the lens, in a non-euclidean spacetime.

In GR, the total deflection angle corresponding to the Schwarzschild solution is given by
\beq
\De\vp_{\text{\tiny GR}}= 2 \abs*{ \int_b^\infty \! \be_{\text{\tiny GR}}\ dr} - \pi
\label{dephiGR}
\eeq
where 
\beq
\be_{\text{\tiny GR}}= \frac{1}{r \sqrt{-1+\frac{\al}{r}-\frac{r^2}{b^2}\left( 1-\frac{\al}{r}\right)    }}\; , \label{betaGR} 
\eeq
$\al=\frac{2 G M}{c^2}$, and $M$ is the total mass of the lens. Equation \eqref{dephiGR} is an elliptic integral. By expanding $\beta$ to first order in $GM/r_0$ before integrating \cite{Weinberg}, we obtain the known result
\beq
\De\vp_{\text{\tiny GR}} \simeq \frac{4 G M}{c^2 b} \; .
\label{dephiGRapprox}
\eeq

For the $\fchi$ theory, we decide not to choose a specific expression of the function in the regime of the
transition between the two limits $\chi\ll \chi_L$ and $\chi \gg \chi_L$ (equation \eqref{systemchi}). Therefore,
we simply split the integral in \eqref{dephi} into the sum of the two contributions on the
scales $r\ll R_M$ and $r\gg R_M$, corresponding to the two curvature regimes. Thus, when the impact parameter $b<R_M$,
the deflection angle is
\beq
\De \vp_{\c} = 2 \abs*{ \int_b^{R_M} \! \be_{\text{\tiny GR}}\ dr + \int_{R_M}^\infty \! \be_{\chi}\ dr} - \pi
\label{dephichi}
\eeq
with
\beq
\be_{\chi}= \frac{  \sqrt{\left(1+2 \ga \right) \left(1+2\ga \log \frac{r}{r_*} \right)}}{ r\sqrt{-1-2\ga \log \left(\frac{r}{r_*}\right)  + \left(\frac{r }{b}\right)^2 \left(1+2\ga \log
   \frac{b}{r_*} \right)}}\; , \label{betaMOND}
\eeq
$\al=\frac{2 G M}{c^2}$, and $\ga=\frac{\left( G M a_0\right)^{1/2}}{c^2}$.
When the impact parameter $b>R_M$, the first integral in equation \eqref{dephichi} drops and the lower
limit of the second integral is $b$; thus we only have the $\chi^{3/2}$ contribution to the deflection angle in the regime $\chi \ll \chi_L$.
We note that, unlike in GR, the metric in \eqref{dephichi} is not asymptotically flat, but, as required by the definition \eqref{dephi},  the spacetime is asymptotically flat, because the scalar curvature $R$ of equation \eqref{ricci} decreases with $r^{-2}$. In addition, the metric component $g_{00}$ only grows logarithmically and $g_{11}$ is constant, and these conditions guarantee that the integral in \eqref{dephichi} converges.

The second integral in equation \eqref{dephichi} also is an elliptic integral, which, in general, cannot be expressed in terms of elementary functions. Expanding $\beta_{\chi}$ to first order in $\ga$ and integrating from an arbitrary radius to infinity, we obtain
\beq
\int_{r}^\infty \beta_\chi(r')\ dr' = 2\frac{ \left( 1 + 2 \ga \right)^{1/2}}{\left(1-\ga \right)} \arctan \left\{ \left[\left(\frac{r}{b}\right)^{2-2\ga}-1 \right]^{-1/2}\right\} - \pi \ . 
\eeq
When $r=b$, the $\be_{\chi}$ contribution to the total deflection angle in $\fchi$ theory is thus
\beq
\int_{b}^\infty \beta_\chi\ dr = 2 \pi \ga =2\pi\frac{\left( G M a_0\right)^{1/2}}{c^2} = 2\pi \left(\frac{v}{c}\right)^2\; ,
\label{xavier}
\eeq
where $v$ is the rotational velocity of disk galaxies appearing in the baryonic Tully-Fisher relation or, in general, the constant circular velocity around the point-mass source.
This equation includes the correct factor 2 that is missing from equation [67] in Mendoza et al. \cite{Mendoza2013}.

\begin{figure}[htbp]
\centering
\includegraphics[width=.6\textwidth]{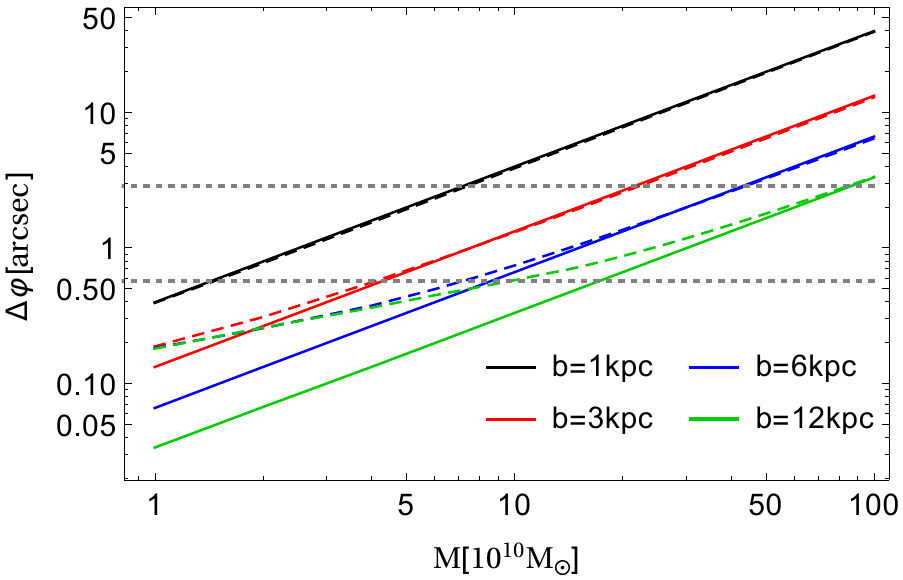}
\caption{Deflection angle $\De\vp$ as a function of lens mass $M$ in the $\fchi$ theory (dashed lines) 
and GR (solid lines). Upper curves correspond to increasing impact parameter $b$. We show curves for $b=[1,3,6,12]$ kpc. The dotted gray lines correspond to the minimum and maximum values of the deflection angles observed in our data sample. }
\label{Anglesgal2}
\end{figure}

\begin{figure}[htbp]
\centering
\includegraphics[width=.6\textwidth]{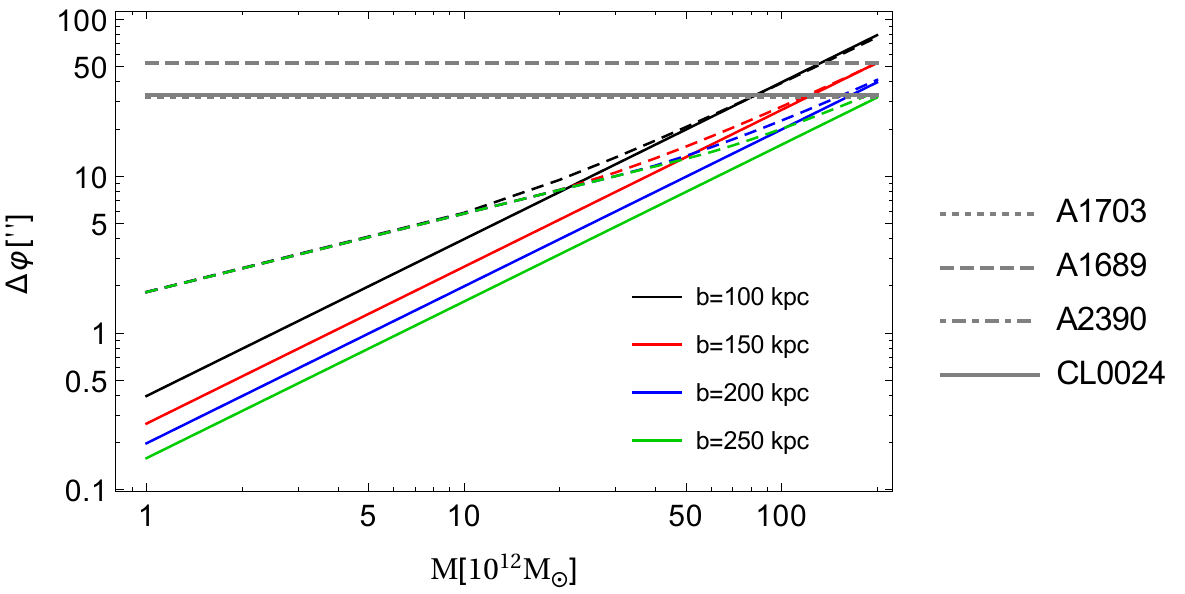}
\caption{Deflection angle $\De\vp$ as a function of lens mass $M$ in the $\fchi$ theory (dashed lines) 
and GR (solid lines). Upper curves correspond to increasing impact parameter $b$. We show curves for $b=[100,150,200,250]$ kpc. The gray lines correspond to the total deflection angles for four clusters of galaxy. }
\label{Anglescl2}
\end{figure}

Figures \ref{Anglesgal2} and \ref{Anglescl2} show how the deflection angle $\De\vp$ [equations \eqref{dephiGR} and \eqref{dephichi}] depends on the lens mass $M$ and the impact parameter $b$ both in GR (solid lines) and the $\fchi$ theory (dashed lines), for typical values of masses and impact parameters of galactic (Figure \ref{Anglesgal2}) and clusters scales (Figure \ref{Anglescl2}). $\De\vp$ increases with $M$ and decreases with $b$ in both theories. 

\begin{figure}[htbp]
\centering
\includegraphics[width=.6\textwidth]{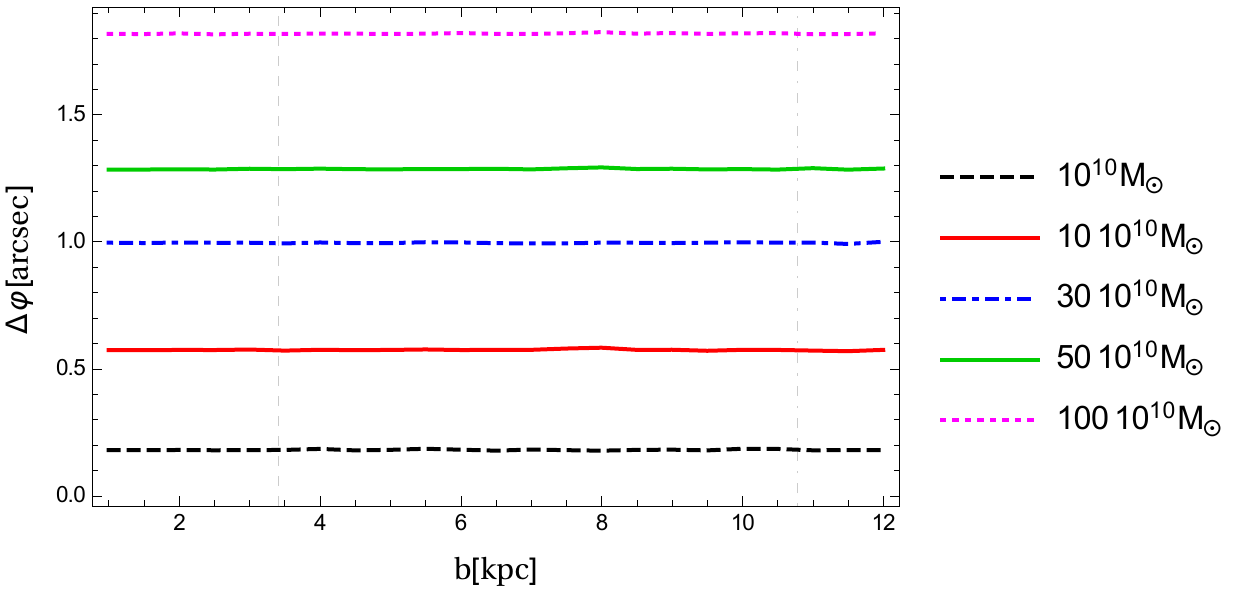}
\caption{Contribution to the deflection angle in the $\fchi$ theory in the weak field limit $\chi\ll \chi_L$, where only \eqref{betaMOND} does contribute to equation \eqref{dephichi}. 
Upper curves correspond to increasing lens masses $M=[1,10,30,50,100]\times 10^{10} M_{\odot}$. The gray dashed and dotdashed lines correspond to the value of $R_M=[3.4, 10.8]$~kpc  for the lens masses $M=[1,10]\times 10^{10} M_{\odot}$, respectively. The $R_M=[18.7, 24.1, 34.1]$~kpc for the larger masses lie outside the plot. }
\label{puremond}
\end{figure}

The dependence on the impact parameter of the deflection angle in the $\fchi$ theory is only due to the GR contribution \eqref{betaGR} to the integral \eqref{dephichi}. In fact, the first integral of \eqref{dephichi} is proportional to the ratio $M/b$, as we can see from equation \eqref{dephiGRapprox}; therefore, in the regime of large $M/b$, the GR contribution dominates the deflection angle. On the other hand, in the regime of small $M/b$, the contribution to the deflection angle mainly comes from  the second term of \eqref{dephichi}, shown in Figure \ref{puremond}: at fixed mass, this term is almost independent of the impact parameter $b$. This independence clearly is expected, because it was imposed by using the order $(v/c)^2$ term [equation \eqref{xavier}] in the derivation of equations \eqref{metricg00a}-\eqref{metricg11b}.

From Figures \ref{Anglesgal2} and \ref{Anglescl2}, we can conclude that the $\fchi$ point mass approximation in the
parameter range explored requires a lens mass smaller than the GR mass to yield the same deflection angle $\De\vp$ with a given impact parameter $b$. This difference is significant only for some values of $M$ and $b$. We will see below whether this difference appears in the ranges of  $M$ and $b$ of the observed astrophysical systems.

\section{Strong-lensing at galactic and cluster scales}
\label{sec:ringssample}
We now apply the point-lens approximation, adopted in our derivation of the deflection angle, to systems where the lensing phenomenon is clearest, namely where the source is as close as possible to the optical axis linking the lens to the observer.
This phenomenon produces the deformation of the image of a background source into a ring, called the Einstein ring, through the lensing of a spherically symmetric foreground object.
In this case, the lens equation yields the following relation between the total deflection angle $\De \vp$ and the observed angular size of the Einstein radius $\th_E$ \cite{SchneiderEhlersFalco} 
\beq
\th_E=\frac{D_{ls}(z)}{D_s(z)} \De \vp \; ,
\label{Einsteinradius}
\eeq
where $D_{ls}$ and $D_{s}$ are the angular diameter distances between the lens and the source and the observer and the source, respectively, in the framework of a Friedmann-Robertson-Walker metric  \cite{Narayan1996, Weinberg}.

Our goal here is to investigate whether the $\fchi$ theory is able to predict the observed angular Einstein radius with the gravitational lensing effect of the luminous matter alone. We are interested in the potential capability of the $\fchi$ theory  to remove the requirement of the presence of dark matter and not in the accurate modeling of the lens system. Therefore, we simply assume that the lens is spherically symmetric and use equation \eqref{dephichi} in equation \eqref{Einsteinradius} to derive the amount of mass required to reproduce the observed angular Einstein radius. For a consistent comparison, we perform the same analysis by assuming GR valid. We refrain from any attempt of accurately modeling the mass distribution of the lens. 

The Sloan Lens ACS Survey (SLACS) \cite{SLACSIX2009} presents a catalog of 85 strong gravitational lens systems whose status is coded by 'A', i.e.~systems with clear and convincing evidence of multiple imaging \cite{SLACSV2008}.
These lenses are in the redshift range $z=[0.063,0.513]$ and  are modeled by singular isothermal ellipticals. Out of these 85 systems, 74 have full or partial Einstein-ring lensed images.
We combine these lenses with a similar sample of 25 systems, in the redshift range $z=[0.3521,0.6591]$, provided by the Boss Emission-Line Lens Survey (BELLS) \cite{BELLS2012} to obtain our final data set of 99 strong-lensing systems.

We now investigate whether the $\fchi$-theory is able to describe the observed phenomenology of our selected sample. 
From the Einstein radius $\theta_E$ and the lens and source redshifts $z$ listed in \cite{SLACSIX2009} and \cite{BELLS2012}, we derive the total deflection angle $\De\vp$ according to \eqref{Einsteinradius}. For the distance estimates, we assume $\Omega_0=0.3$, $\Omega_{\Lambda 0}=0.7$, and $H_0=(67.8 \pm 0.9)\text{Km s}^{-1}\text{Mpc}^{-1}$ \cite{PlanckCosmo2015}.
The gray lines in Figure \ref{Anglesgal2} show the lower and upper bounds of the total deflection angles for our sample: the observed $\De\vp$ range corresponds to a region where the two theories are almost indistinguishable; in other words, the two theories reproduce the same deflection angle with the same amount of matter.
With our analysis, we find that the average mass-to-light ratio  of the lenses is $\sim 2\ M_\odot/L_\odot$, within a factor of 1.3 of the mass-to-light ratios derived from the SLACS and BELLS collaborations that adopt a lens mass model more sophisticated than our simple point-like lens. These mass-to-light ratios are roughly consistent with the ratios expected for pure old stellar populations, and we can conclude that, for these lens galaxies, $\fchi$ might require the same amount of dark matter, albeit small or even zero, that is required by GR.

We thus turn to strong lensing in galaxy clusters where the observed partial Einstein rings or arclets, interpreted within the GR framework, imply a large amount of dark matter. A few notable examples are A2390 and CL0024+17. In these two clusters, the gas mass fraction, that is a reasonable rough estimate of the ratio between the amounts of ordinary
and total matter within the cluster, is, respectively, $f_{\text{gas}} \sim 0.21$ at $0.9$ Mpc \cite{Allen2001}, and $f_{\text{gas}} \sim 0.14$ at $R_{200}=2.73$~Mpc \cite{Ota2004}.
These gas fractions suggest that at least 80 percent of the cluster mass within these radii should be dark. Therefore, the two theories should exhibit a rather different behaviour in clusters of galaxies. 

We model four clusters with our point-like lens approximation in GR and $\fchi$: CL0024, A1689, A1703, and A2390. The third column of Table \ref{Table1} lists the GR masses within the Einstein radius $r_{\rm E}$ derived with accurate modelings of the lenses as reported in the literature; the fourth column reports our GR mass: the largest discrepancy between the two estimates occurs for CL0024 and amounts to 28.4 percent.  
The cluster masses within $r_{\rm E}$ we derive in $\fchi$ are listed in the fifth column of Table \ref{Table1} and are within a few percent of our GR masses: the largest differences is 7.5 percent, again for CL0024. This agreement is consistent with the gray lines in Figure \ref{Anglescl2}, which show the $\De\vp$ for the four clusters. The last column of Table \ref{Table1} lists the gas fraction $f_{\rm gas}$ within  $r_{\rm E}$ estimated from X-ray observations within the GR framework as reported in the literature. For A1703, we list the mass-to-light ratio in the $R$ band \cite{Medezinski2010}. For A1689, A1703, and A2390, we report the values derived by visual inspection of the profiles reported in the literature. The agreement between our GR and $\fchi$ cluster masses and the values $f_{\rm gas}\ll 1$ (or $M/L\gg $ a few) unambiguously show that 
$\fchi$ requires the same amount of dark matter as GR to describe the gravitational strong  lensing phenomenology in the core of galaxy clusters.

 
\begin{table}
\begin{center}
\caption{The Einstein radius  $r_{\text{\tiny E}}$ and the enclosed GR masses as reported in the literature and the GR and $\fchi$ masses estimated with our point-like approximation; the last column lists the enclosed gas fraction. For A1689 and A2390, we estimated the gas fraction at the Einstein radius from the plots in \cite{Lemze2008} and \cite{Allen2001}, respectively. For A1730, we list the mass-to-light ratio reported in \cite{Medezinski2010}.}
\begin{tabular}{cccccc}
\toprule
 & $r_{\text{\tiny E}} [\text{kpc}]$ & $M_{\text{\tiny GR}}^{\text{ lit}} [10^{14} \text{M}_\odot]$ & $M_{\text{\tiny GR}} [10^{14} \text{M}_\odot]$ & $M_{\text{\tiny $f(\chi)$}} [10^{14} \text{M}_\odot]$ & $f_{\text{gas}}$\\
\midrule
CL0024 & 191 & 2.212 $\pm$ 0.003 \cite{Ota2004}  & 1.584 & 1.495 & 0.28$^{\text{\tiny $+$0.29}}_{\text{\tiny $-$0.23}}$ \\
A1689  & 140 & 1.85 $\pm$ 0.26 \cite{Limousin2006} & 1.661 & 1.663 & $\sim$ 0.04 \\
A1703  & 133 & 1.21 $\pm$ 0.10 \cite{Zitrin2010}& 1.059 & 1.035 & $\sim 240 M_\odot/L_\odot$\\
A2390  & 144 & 1.6  $\pm$ 0.2 \cite{Allen2001}& 1.186 & 1.155 & $\sim$ 0.15 \\
\bottomrule
\end{tabular}
\label{Table1}
\end{center}
\end{table}


\section{Conclusions and perspectives}
\label{sec:conlusions}

Here we have estimated the light deflection  as a function of lens mass and impact parameter in the $\fchi$ theory of gravity,
an $f(R)$ gravity where the action of the gravitational field depends on the source mass; $\fchi$ has been suggested to reproduce the gravitational lensing phenomenology without the need of dark matter \cite{Mendoza2013,Bernal2015}. Reproducing
this phenomenology with baryonic matter alone and without any additional scalar or vector fields would be an unprecedented results. 

For our computation, we have used a metric solution accurate to second order in $v/c$ for a point-mass lens.
We analyse galaxy lensing systems from SLACS and BELLS and a sample of four galaxy clusters in the strong lensing regime. Although for some combinations of lens mass $M$ and impact parameter $b$, the light deflection in $\fchi$ is larger than in GR,  the two gravity theories are indistinguishable for most lens masses and impact parameters of astrophysical interest on galaxy and galaxy cluster scales.
Our result is particularly important for the inner regions of  clusters,  where, in the GR framework, the mass associated to stars and the intergalactic medium accounts for less than roughly 20 percent of the total mass of the cluster within the same region.

Our approximated metric solution of the $\fchi=\chi^{3/2}$ theory thus appears to be unable to describe the gravitational lensing phenomenology without assuming the presence of dark matter. However, the radial component \eqref{metricg11a}-\eqref{metricg11b} of our metric solution has been derived by imposing the validity of the baryonic Tully-Fisher relation and the independence of the light deflection angle from the impact parameter. We have not yet investigated whether additional solutions to $\fchi=\chi^{3/2}$ exist that meet the same requirements and can describe the inner region of clusters of galaxies without the aid of dark matter.  

Our results appear to be at odds with Bernal et al. \cite{Bernal2015} claim that the  $\fchi=\chi^{3/2}$ theory
is able to describe the mass profiles of a sample of 12 X-ray clusters of galaxies without the need of dark matter. This ability would be relevant because MoND, whose successes on galactic scales are shared by the $\fchi$ theory, is known to fail on cluster scales. A possible reason for this apparent discrepancy is in the order of the metric expansion in the two analyses: in our lensing investigation a second order expansion suffices, whereas the dynamics of galaxies in clusters require a fourth order expansion; therefore the former analysis has one free parameter, that we call $r_*$, whereas in the latter, Bernal et al. have three free parameters, $r_*$, $A$, and $B$, that vary by more than a factor of $10^4$ ($r_*$), $20$ ($A$), $2$ ($B$), to fit the observed X-ray temperature profiles. In our lensing analysis, the only free parameter $r_*$ has a much smaller impact on the amount of light deflection: varying $r_*$ by a factor of $10^6$ induces a relative variation smaller than a factor of $10^{-3}$ in the deflection angle. An alternative solution to the discrepancy between Bernal et al.'s results and ours might be that in $\fchi$ the deflection angle sensitively depends on the accurate modeling of the mass distribution within the lens: if this is the case, a model more sophisticated than our simple point-like lenses might substantially change our conclusion and remove the need for dark matter in our galaxy cluster lens systems. 
However, it appears to be unlikely that this features should apply to $\fchi$ and not to GR; in fact, the discrepancies between the cluster core masses derived in GR with our point-like lens and the mass reported in the literature (Table \ref{Table1}) are not large enough to remove the need of dark matter, and we might in principle expect the same behaviour in $\fchi$.
On the contrary, a more accurate mass modeling might be worth investigating if tests of $\fchi$ in parameter ranges where it yields distinct results from GR, namely the weak lensing regime, turn out to be successful.

Finally, here we have tested the metric \cref{metricg00a,metricg11b} that are solutions of the $\fchi$ theory as a power law, with $\fchi=\chi^{3/2}$: this function is the only power law that shares the successes of MoND on galactic scales \cite{Bernal2011,Mendoza2012}. It remains to be seen whether a
more complicated functional form of $\fchi$ can both mimic MoND on small scales and the presence of dark matter on 
cluster and larger scales. 


\section*{Acknowledgments}
MC and AD acknowledge partial support from INFN grant InDark, and the grant PRIN 2012 'Fisica Astroparticellare Teorica' of the Italian Ministry of University and Research. LF acknowledges the INFN grant QGSKY, the local research project {\it  Metodi Geometrici in Fisica Matematica e Applicazioni (2015)} of Dipartimento di Matematica of University of Torino (Italy), and the grant INdAM-GNFM. The work of XH was supported in part by DGAPA-UNAM PAPIIT IN-100814 and CONACyT. This article is based upon work from COST Action (CA15117 CANTATA), supported by COST (European Cooperation in Science and Technology).

\begin{appendices}

\section{Geodesic trajectories}
\label{appendixA}

We derive the time component of the metric, i.e. the gravitational potential, from the study of the timelike geodesics in a spherically symmetric spacetime
\beq
g=A(r) dt^2 -\frac{dr^2}{B(r)}-r^2 d\th^2 - r^2 \sin^2 (\th) d \phi^2 \; .
\label{symmg}
\eeq
By limiting the geodesics to the plane $\th=\pi/2$, the Lagrangian corresponding to the metric \eqref{symmg} is
\beq
L=\sqrt{A(t')^2-\frac{(r')^2}{B} -r^2(\phi')^2} ds=\sqrt{A-\frac{\dot{r}^2}{B} -r^2\dot{\phi}^2} dt \; ,
\eeq
where the apostrophe ($'$) indicates the derivative with respect to the generic parameter $s$, and the dot ($\dot{}$) the derivative with respect to the time coordinate. Since the coordinates $(t,\phi)$ are cyclic, we have the two first integrals
\beq
J=\frac{-r^2 \dot{\phi}}{\sqrt{A-\frac{\dot{r}^2}{B}, -r^2\dot{\phi}^2}} \qquad E=\frac{A}{\sqrt{A-\frac{\dot{r}^2}{B} -r^2\dot{\phi}^2}}
\eeq
and the velocities
\beq
\phidot^2=\frac{J^2}{r^4}\frac{A^2}{E^2}, \qquad \rdot^2=AB - \frac{A^2B}{r^2E^2}\left(r^2 + J^2 \right) \; .
\label{velocities}
\eeq
The circular motions correspond to the initial conditions
\beq
r(0)=r_0, \quad \rdot(0)=0, \quad \phi(0)=0, \quad r_0\phidot(0)=v_0 \; ,
\eeq
corresponding to the constants
\beq
J=\frac{-r_0 v_0}{\sqrt{A_0 -v_0^2}}, \qquad E=\frac{A_0}{\sqrt{A_0 -v_0^2}} \; .
\label{firstintegralconst}
\eeq
Once we substitute these constants into the second equation of \eqref{velocities}, we obtain 
\beq
\rdot^2=AB \left[1-\frac{A}{A_0}+\frac{A}{A_0^2}\frac{v_0^2}{r^2}\left( r^2-r_0^2\right) \right] = \Phi(r;r_0,v_0) \; .
\label{Phi}
\eeq
Imposing the motion to be circular requires the following conditions on equation \eqref{Phi} and its derivative
\beq
\Phi(r_0;r_0,v_0)=0, \qquad \Phi'(r_0;r_0,v_0)=0 \; .
\eeq
The first relation is trivially verified, while the second provides the following expression
\beq
\Phi'(r_0;r_0,v_0)=B_0 \left(-A_0'+2\frac{v^2}{r_0}\right)
\eeq
that is zero if and only if we set
\beq
v^2(r)=\frac{1}{2} A'(r) r \; .
\label{correctV}
\eeq
Equation \eqref{correctV} is the relation between the particle velocity and the time component of the metric that describes the circular motion of the particle around the point-mass source. Equation \eqref{correctV} can be used to constrain the time component of a metric describing a flat rotational curve of a disk galaxy. By using the Tully-Fisher relation with
a constant rotational velocity, $M_b= v^4/(Ga_0)={\rm const}$, with $a_0=1.2\times 10^{-10}$~m~s$^{-2}$, we can integrate $A(r)=-g_{00}$ from \eqref{correctV} to obtain
\beq
g_{00}=-1-2\frac{(G M_b a_0)^{1/2}}{c^2} \log\left(\frac{r}{r_*}\right)
\label{metrict}
\eeq
\beq
A(r)=1+2\frac{(G M_b a_0)^{1/2}}{c^2} \log\left(\frac{r}{r_*}\right) \; .
\eeq
We finally derive the gravitational potential in the Newtonian limit 
\beq
\phi(r)=(G M_b a_0)^{1/2} \log\left(\frac{r}{r_*}\right) \; ,
\eeq
which provides the correct sign for an attractive acceleration. 

\end{appendices}

\bibliographystyle{JHEP}
\bibliography{Xavierbib}

\end{document}